\documentclass[journal]{IEEEtran}

\bstctlcite{IEEEexample:BSTcontrol}

\normalsize

\usepackage{multirow}

\usepackage{amssymb,amsthm}

\usepackage{cite}

\usepackage[cmex10]{amsmath}

\usepackage{fixltx2e}

\usepackage{stfloats}

\usepackage{graphicx}

\begin{document}
%
\title{Ultra Dense Networks: The New Wireless Frontier for Enabling 5G Access}

%
%
%

\author{Antonis~G.~Gotsis,~Stelios~Stefanatos,~and~Angeliki~Alexiou
\thanks{The authors are with the Department
of Digital Systems, University of Piraeus, Greece. email: \{agotsis, sstefanatos, alexiou\}@unipi.gr.}
\thanks{
This work has been performed in the context of the “ARTCOMP PE7(396) Advanced Radio Access Techniques for Next Generation Cellular NetwOrks: The Multi-Site Coordination Paradigm” research project, implemented within the framework of Operational Program “Education and Lifelong earning”, co-financed by the European Social Fund (ESF) and the Greek State.
}
}


\markboth{}%
{}

\maketitle

\begin{abstract}
The extreme traffic load that future wireless networks are expected to accommodate requires a re-thinking of the system design. Initial estimations indicate that, different from the evolutionary path of previous cellular generations that was based on spectral efficiency improvements, the most substantial amount of future system performance gains will be obtained by means of network infrastructure densification. By increasing the density of operator-deployed infrastructure elements, along with incorporation of user-deployed access nodes and mobile user devices acting as ``infrastructure prosumers'', it is expected that having one or more access nodes exclusively dedicated to each user will become feasible, introducing the ultra dense network (UDN) paradigm. Although it is clear that UDNs are able to take advantage of the significant benefits provided by proximal transmissions and increased spatial reuse of system resources, at the same time, large node density and irregular deployment introduce new challenges, mainly due to the interference environment characteristics that are vastly different from previous cellular deployments. This article attempts to provide insights on fundamental issues related to UDN deployment, such as determining the infrastructure density required to support given traffic load requirements and the benefits of network-wise coordination, demonstrating the potential of UDNs for 5G wireless networks.
\end{abstract}


%
\IEEEpeerreviewmaketitle

\section{Introduction}
``\textit{From the smallest personal items to the largest continents, everything, everywhere will be digitally connected, and responsive to our wants and likes}'', is a typical vision for the future of communication networks as stated in “The World in 2025: 10 Predictions of Innovation'', (Thomson Reuters). In particular, wireless communications are expected to dominate everything, everywhere, mainly empowered by revolutionary 5th generation (5G) technologies: user-centric, cell-less architectures, massive spatial processing, tactile response times and big data processing, to name a few. Towards this end, the quest for 5G has already begun \cite{AnBu14,LiNi14}: international industry-academia consortia (METIS2020, IWPC, 5G Korea), public-private partnership associations (5G-PPP), regulatory authorities (ITU-R WP5D) and operators fora (NGMN), have recently focused their attention on setting the requirements, conceptualizing, and exploring the enabling technologies for 5G \cite{OsBo14}.

Wireless evolution has come a long way over the last 30 years, shifting the focus of innovation from components and transceiver technologies (e.g., turbo coding, MIMO) to resource allocation and network management (e.g., CDMA, OFDMA, CoMP) and, finally, to user-centric and device-driven communications towards realizing the concept of ``internet of everything''. Accelerated by the dramatic impact of applications of the kinds of social networking, the center of gravity for 5G will be shifted further to application-driven connectivity (transparently deployed over technologies, infrastructures, users and devices), which will be mainly enabled by means of dense access/serving node deployments and exploitation of proximal communications, concepts that are typically referred to as network densification \cite{BhJu14}. In contrast to previous cellular evolution steps, which were mainly driven by increasing system bandwidth and improving (multi-user) spectral efficiency, network densification is expected to be the cornerstone of future wireless networks. Even though network densification was also considered in past cellular generations, it was used only as an ``add-on'' measure for locally coping with heavily loaded parts of the system and not as a fundamental, large-scale design/system concept.

\begin{figure}
\centering
\resizebox{8.5cm}{!}{\includegraphics{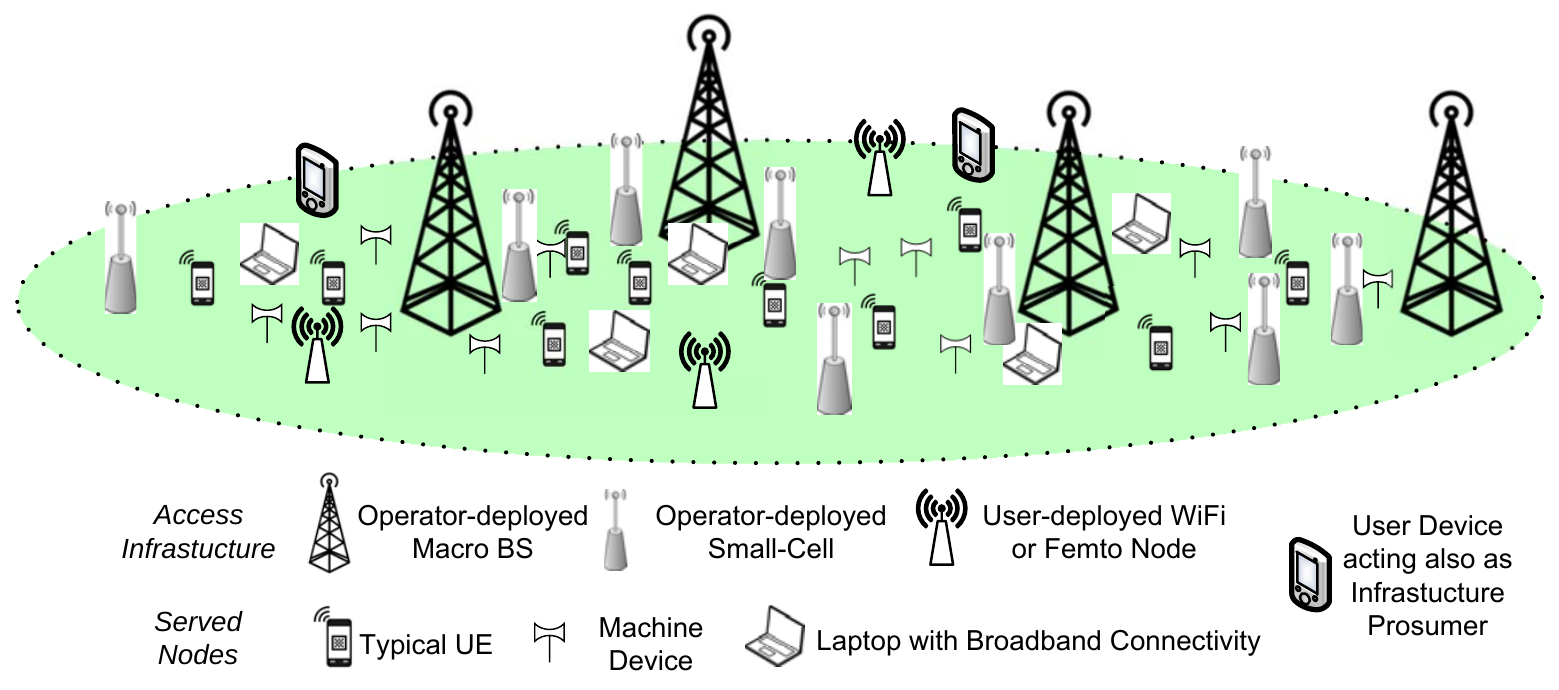}}
\caption{An ultra-dense network infrastructure composed by operator- and user-deployed heterogeneous serving access nodes, multiple types of user and machine served nodes, and disruptive devices acting as prosumers. For a given area the number of serving and user nodes are of the same order.}
\label{Fig1}
\end{figure}

Towards realizing the ambitious goals set for 5G, the density of access/serving nodes is expected to increase up to the point where it is comparable to or even surpass the (also increasing) density of user equipments (UEs) \cite{MoPa14}, thus introducing the ultra-dense network (UDN) paradigm. Realization of such a disruptive network topology, with respect to the evolution path of previous cellular generations, will be achieved by utilizing, in addition to densified traditional access nodes (ANs) such as small cells, user deployed ANs (e.g., WiFi, femtocells), as well as ``infrastructure prosumer'' UEs, i.e., (mobile) devices with computational and storage capabilities allowing them to act as infrastructure ANs (see also Figure \ref{Fig1}).

Clearly, a successful UDN deployment is not an easy task due to the challenging interference landscape introduced by the dense and, to a large extent, irregular AN deployment with transmissions under a universal frequency reuse scheme. In this sense, UDNs can be viewed as the new wireless frontier that can be made a reality only through re-thinking of current cellular practices in terms of: i) underlying technology enablers, namely, air-interface, including optimized frame structures and associated overhead signaling, resources management, and backhaul design \cite{MoPa14}; and ii) implementation implications, such as business modeling, cost/complexity, backward compatibility, and market uptake issues. As typical in any engineering application, in order to successfully and efficiently address these issues, a clear understanding of UDN fundamental performance limits, obtained using appropriate mathematical models, is of utmost importance. Towards this end, two relevant broad research challenges can be identified that are not adequately addressed to date by the research community:
\begin{enumerate}
\item How should AN density scale in order to accommodate increasing traffic load requirements due to increasing UE density and/or increasing UE rate requests?
\item Is densification alone sufficient to accommodate future network requirements? In that respect, does network-wise coordination offer significant performance gains that would justify its incorporation in UDNs?
\end{enumerate}
By shedding some light on the above aspects, this article aims at revealing some of the critical deployment design factors for UDNs and attempt a qualitative and quantitative assessment of their potential.

The rest of the article is structured as follows. In the next section we present a brief overview of the latest and ongoing work on the topic. Then, after positioning UDNs in 5G network evolution, we introduce the two aforementioned UDN research challenges. In the next two sections we investigate the two challenges, by providing the necessary background, latest contributions and indicative UDN performance evaluation results. In the last section we conclude the article and state potential future research directions.

\section{Recent Work on Ultra-Dense Networks}
The concept of adding smaller footprint cells within macro-cell deployments as a means of offloading traffic in hotspot localized areas, known as Heterogeneous Networks (HetNets), has set the scene for the UDN paradigm. In \cite{HwSo13}, a view of ultra dense HetNets, towards realizing the so-called 1000$\times$ capacity challenge, was presented. It was shown through theoretical analysis and system-level simulations that such a challenging target could be achieved by combining network densification, spectrum increase, and advanced radio access technologies and concepts (such as Massive MIMO, CoMP, D2D), with densification contributing the most to this increase. 

In \cite{YuVa15}, spectral and energy efficiency aspects of UDNs considering different deployment strategies where investigated, involving heterogeneous access node types, namely, conventional macro-cells, indoor femto-cells, and outdoor distributed antenna systems (DASs). A scheme where dynamic outdoor DASs are utilized for providing outdoor capacity on demand, with femto-cells dedicated for serving indoor and, opportunistically, outdoor traffic, was proposed. 

Another research direction involves the operation of UDNs in mm-Wave bands, due to the availability of vast amount of frequency resources compared to traditional bands, allowing to realize 5G use cases, such as “amazingly fast” multi-Gbps speeds \cite{OsBo14}. In \cite{BaIr15}, this new system concept was introduced, and the key design challenges were presented. In particular, potential OFDM-based physical layer structures were proposed for realizing UDN mm-Wave PHY, the importance of beamforming was highlighted, and self-backhauling along with spectrum sharing were proposed as major technology enablers.

The importance of the UDN paradigm for future networks has also already been recognized by the industry, as evidenced by the activities of two major standardization bodies, 3GPP and IEEE 802.11 working group.  In 3GPP, Releases 12 and onwards promise ``taking the use of small cells and network densification to a whole new level'' \cite{3GPP_36872} by capitalizing on the gains of ultra-dense networks, whereas in the WLAN community, a recent project called 802.11ax (formerly known as High-Efficiency Wi-Fi or simply ``HEW'') aims at delivering a new standard that promises significant enhancements at highly dense deployment scenarios. 

With respect to recent R\&D activities, METIS2020, which constitutes the European pre-5G flagship initiative, considers UDN a major topic for facilitating the future 5G system concept \cite{OsBo14}. Early system evaluation results point out that heavy infrastructure densification (up to the point that a single access node serves at most three user devices) is a prerequisite for sustaining 1000 times greater mobile data volumes than of today, a key requirement for 5G wireless access.

\section{Ultra Dense Networks: Objectives, Progress, Analysis and Design Methodologies}

\subsection{Fundamental Challenges of Ultra Network Densification}
Cellular infrastructure densification is a concept that has been applied as early as in 2G voice-oriented systems, mainly as a tool for locally enhancing performance of an already deployed system. In particular, cell splitting and sectorization, supported by careful frequency planning were employed in order to increase sustained voice capacity in areas of the network experiencing heavy traffic. Regarding future networks, (ultra) network densification has a fundamentally different and more important role. In particular, network densification offers the advantage of proximal communications, which, in turn, provide the means towards fulfilling the following critical communication principles:
\begin{itemize}
\item Extreme, user-centric, spatial reuse of system bandwidth, by partitioning the space to (arbitrarily) small cells, up to the point where the traditional notion of cell is no longer relevant and each UE is served in principle by one (or more) AN(s) exclusively dedicated to that UE;
\item Improvement of propagation conditions, by bringing the access infrastructure closer to the user.
\end{itemize}

It is clear that the proximal communications advantage offered by UDNs translates to an improvement in the received useful signal power from the serving AN(s). However, for the same reasons, the signal power from interfering transmissions is also increased. It is therefore necessary to understand how the combination of these two effects affects system performance. What makes this task non-trivial compared to similar studies performed for legacy cellular deployments are the unique characteristics of UDNs, namely, irregular deployment and elimination of the conventional notion of cell.

\begin{figure}
\centering
\resizebox{7.5cm}{!}{\includegraphics{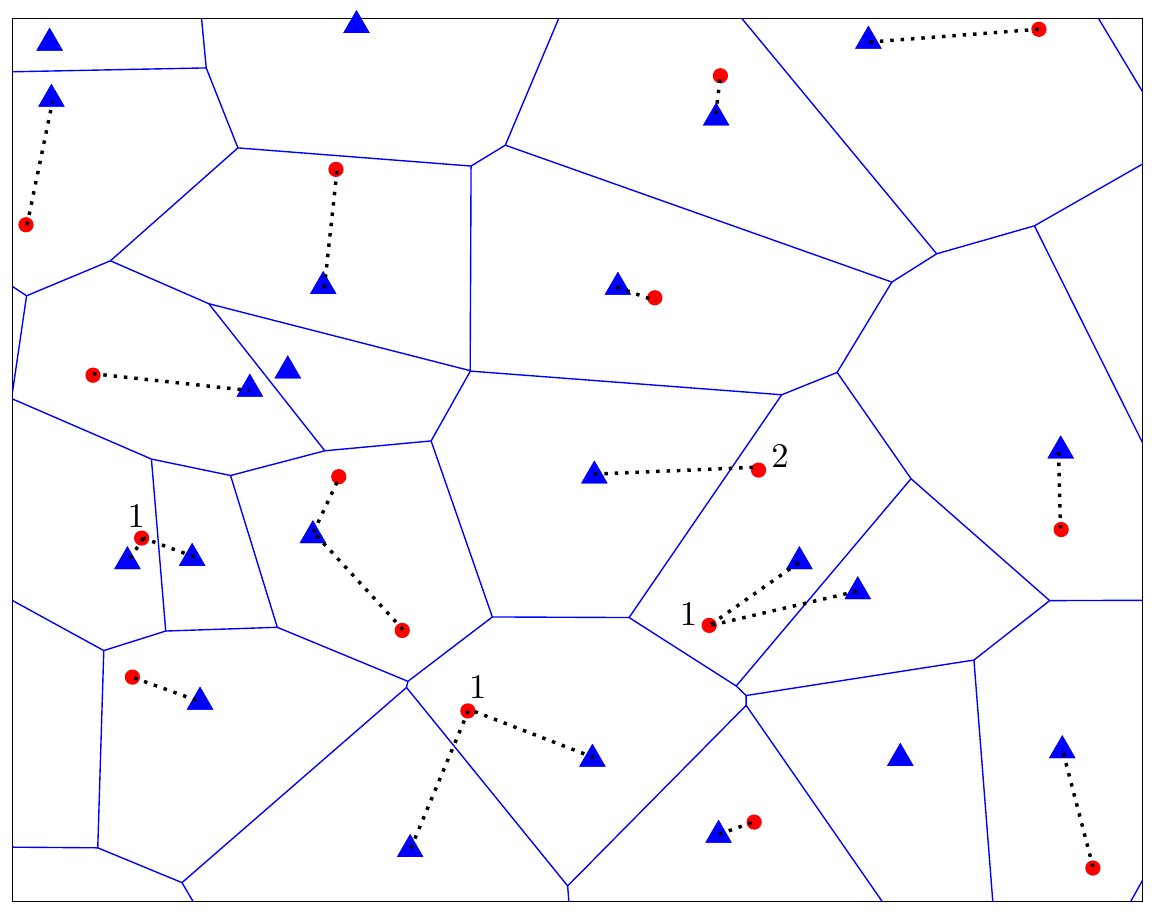}}
\caption{Irregular AN deployment in UDNs within an area previously corresponding to a single cell of a conventional cellular network. Triangles correspond to ANs and circles to UEs, with (possible) AN-UE associations depicted by dotted lines. Note that the UDN offers the flexibility of having more than one ANs exclusively serving a single UE (UEs labeled ``1'') or a UE associated with an AN that is not its closest but having to serve less traffic (UE labeled ``2'').}
\label{Fig2}
\end{figure}

The irregular deployment of ANs is due to practical limitations of infrastructure deployment, opportunistic introduction of low-cost ANs, and exploitation of user (mobile) devices as content providers. This results in a completely different interference environment than the one experienced in previous cellular generations, e.g., where interferers may be positioned arbitrarily close to a receiver. In addition, the UDN topology offers new possibilities on how a UE may be served by the network, going beyond the conventional, cell-based scheme. In a UDN deployment, many ANs may be tagged as potential  (exclusive) servers to a particular UE, resulting in new AN-UE association patterns, e.g., a UE may choose not to connect to the nearest AN when the latter is heavily loaded (see Figure \ref{Fig2}).  These special attributes of UDNs open new research questions, which can be broadly divided into two frameworks:
\begin{itemize}
\item analytical system modeling/characterization, and 
\item algorithmic radio resource management.
\end{itemize}

\subsection{UDN Analysis}
Towards understanding the impact of densification under irregular (random) AN placement, the seminal work of \cite{AnBa11} leveraged stochastic geometry for analyzing the downlink SIR distribution experienced in an infinite area network with AN positions modeled as a realization of a homogeneous Poisson point process (HPPP), uniquely characterized by its density. The authors showed that in an ever-densified interference-limited network, where UEs associate with their closest AN and no coordination among transmissions is applied, interference and useful signal power increase at the same rate, hence the achieved signal-to-interference (SIR) levels are preserved. It can therefore be concluded that, from a system perspective, network densification indeed enables extreme spatial reuse of system bandwidth, with an increase of AN density directly translated into an increase of total system capacity. 

However,  the system perspective of \cite{AnBa11} as well as of relevant works that followed \cite{ElHo13}, ignored the impact of multiple-access/resources allocation on user rate; the latter is of great importance towards characterizing the perceived rate performance levels on a user-centric future network deployment. In this respect, new tractable analytical modeling approaches focusing on user-centric performance metrics and covering various density regions are necessary for thoroughly characterizing the impact of densification. A relevant study and its application to several envisioned UDN setups will be presented in Section IV.

\subsection{UDN Resources Management Optimization}
Although analytical assessment approaches are able to provide insights on the fundamentals of UDNs operation, their tractability heavily depends on the adoption of (over)simplifying assumptions and abstractions. Therefore, detailed system-level simulation models are necessary in order to capture the effect of, e.g., realistic traffic models and propagation conditions, and investigate the potential of sophisticated radio resource management algorithmic approaches.

In this context, a second line of works has recently emerged in the literature, towards estimating the densification requirements for a set of potential 5G traffic load and QoS targets \cite{HwSo13,HuSa13,GeLe13}. The common conclusion  is that the ambitious set of 5G capacity objectives requires one to two orders of magnitude greater infrastructure density than today, i.e., validate the necessity of UDNs. An extension of such studies is to consider performance optimization aspects, achieved through intelligent, network-wise coordination mechanisms. For example, \cite{HuQi14} considered a system framework for optimizing the spectrum and energy efficiency in multi-layer heterogeneous networks through employing joint interference management, mobile association and cooperative dynamic resource allocation. However, in this paper ultra-dense deployments of low-power nodes were not considered.

Investigation of coordination is of great importance as it has the potential to control interference and exploit the proximity of a UE to many ANs, a system attribute inherent in UDNs and not present in previous wireless generations. However, it is not clear whether incorporation of sophisticated coordination on top of a UDN is indeed necessary since it may be argued that the proximity gains offered by the dense deployment alone are sufficient. In Section V, a relevant network optimization framework will be discussed and applied in the UDN context identifying the benefits of coordination in terms of UE rate and infrastructure densification levels.

\section{Densification Requirements and Scaling laws for UDNs}
Understanding if and how the user-centric requirements of the future network can be provided by infrastructure densification is clearly of critical importance. Towards this end, a mathematical system model is required that incorporates and provides design insights on the following parameters:
\begin{itemize}
\item infrastructure densification level, reflected by the density of ANs, ${\lambda _{\text{AN}}}$ (in number of ANs per unit area),
\item network load, reflected by the density of UEs, ${\lambda _{\text{UE}}}$  (in number of UEs per unit area),
\item per-user rate (in bps/Hz).
\end{itemize}

These parameters were jointly considered in \cite{StAl14tcom} where the downlink UE rate distribution was examined for a randomly deployed, interference limited wireless network as the one considered in \cite{AnBa11}, taking into account the effect of resource sharing using an OFDMA-like multiple-access scheme and assuming an SIR threshold $\theta_0$ below which no service can be provided (e.g., due to synchronization requirements). Each single-antenna UE is served by its closest single-antenna AN, assumed to have full buffers, with ANs transmitting with a fixed power over active subchannels without any coordination among them.

Under typical assumptions employed in stochastic geometry-based works, namely, UEs and ANs distributed as independent HPPPs, log-distance path-loss model with exponent $\alpha$, Rayleigh fading, and conditioning on the existence of a (typical) UE with arbitrary position, a closed form expression for the typical UE rate distribution can be obtained as a function of  ${\lambda _{{\text{AN}}}},{\lambda _{{\text{UE}}}},{\theta _0},\alpha$, and the number of OFDMA subchannels \cite{StAl14tcom}. Interestingly, dependence on the AN and UE densities is only in terms of their ratio, which we define as the densification ratio $\tau  = {{{\lambda _{{\text{AN}}}}} \mathord{\left/
 {\vphantom {{{\lambda _{{\text{AN}}}}} {{\lambda _{{\text{UE}}}}}}} \right.
 \kern-\nulldelimiterspace} {{\lambda _{{\text{UE}}}}}}$. Densification ratio can be conveniently utilized as a metric for differentiating between “traditional”, sparse infrastructure deployments corresponding to small $\tau$  (typically, $\tau  \ll 1$), and (future) ultra-dense deployments for which  may even exceed unity.
 
\begin{figure}[h]
\centering
\resizebox{7.5 cm}{!}{\includegraphics{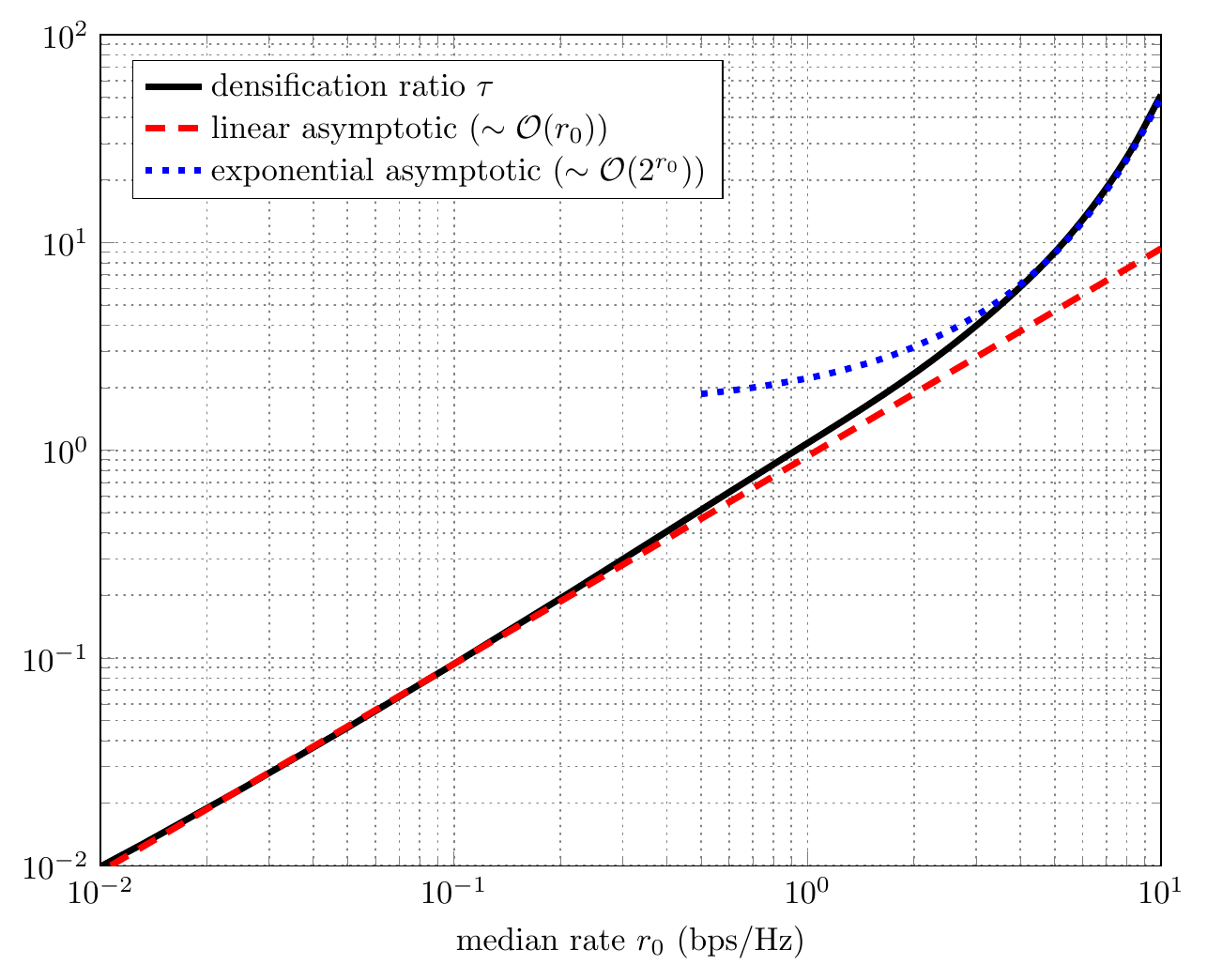}}
\caption{Minimum required densification ratio as a function of target UE median rate ($\alpha = 4$, $\theta_0=-6$ dB). }
\label{Fig3}
\end{figure}

The rate distribution expression allows for efficient optimization of system design parameters such as densification ratio and number of subchannels. As an example case of particular interest in the deployment of UDNs, Figure \ref{Fig3} illustrates the minimum $\tau$ required to provide the typical UE a target median rate $r_0$. It can be shown \cite{StAl14tcom} that the minimum densification ratio scales linearly with $r_0$, when  ${r_0} \to 0$, whereas it scales exponentially with $r_0$ when ${r_0} \to \infty $. These asymptotics are also depicted in Figure \ref{Fig3} where it can be seen that they are quite accurate even for moderately large and small values of $r_0$. This behavior of $\tau$ has a significant impact on future network design as it may imply unrealistically high densification ratios when (excessively) large UE rate levels are required. Therefore, a practical limit on the benefits of (ultra) densification does exist in this respect, which can only be overcome by other means such as spectrum increase and multiple-antenna transmission techniques.

\begin{figure}[h]
\centering
\resizebox{7.5 cm}{!}{\includegraphics{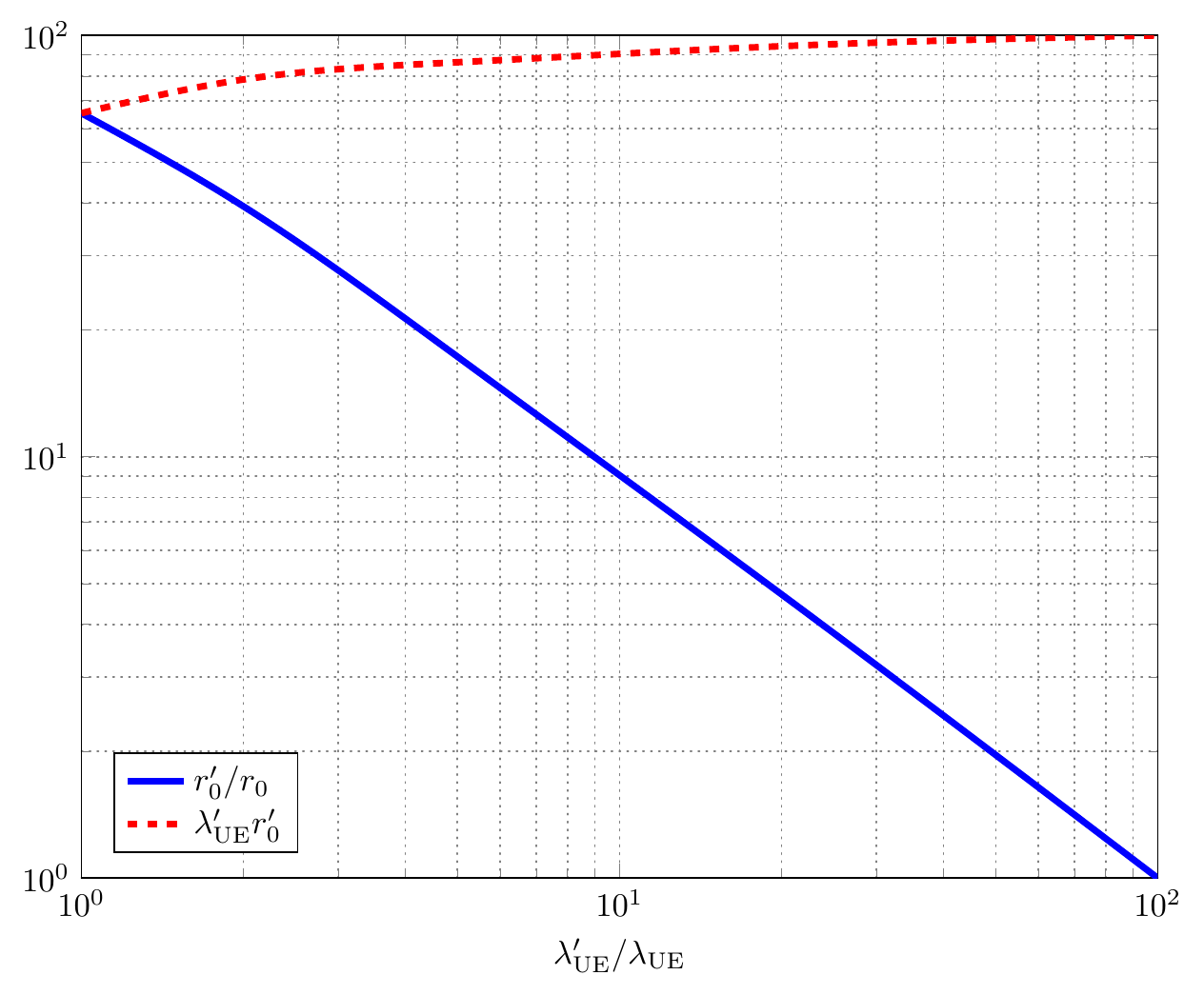}}
\caption{Exploiting AN densification in terms of increasing UE density and/or per UE median rate (initial network state: $\lambda_\textrm{AN} = \lambda_\textrm{UE}/20$, final network state: $\lambda_\textrm{AN}' = 100  \lambda_\textrm{AN}$).}
\label{Fig4}
\end{figure}

Another important issue for UDNs is how the network capacity increase offered by infrastructure densification, i.e., increase of  ${\lambda _{\text{AN}}}$, should be exploited. As described above, with a fixed  ${\lambda _{\text{UE}}}$, increasing ${\lambda _{\text{AN}}}$ and, therefore  $\tau$, can be translated into an increase in achieved rate (in a statistical sense). This constitutes one approach of exploiting densification that is relevant, e.g., for supporting new applications/services with high rate requirements. The other approach is to consider a fixed UE rate requirement and increase ${\lambda _{\text{UE}}}$ proportionally to ${\lambda _{\text{AN}}}$ so as to keep $\tau$ equal to its minimum value required to support the particular rate level. This approach results in a system capable of serving (much) more UEs without per-UE performance degradation, which is relevant for, e.g., supporting machine-to-machine communications, where many low-rate devices are introduced to the system. 

Clearly, there are scenarios in-between these two extremes and Figure \ref{Fig4} considers these in an example case where, starting from a baseline, sparse cellular scenario with ${\lambda _{\text{AN}}} = 5$ serving a network load ${\lambda _{\text{UE}}}=100$, the density of ANs is increased to a new value ${\lambda _{{\text{AN}}}}^\prime  = 100{\lambda _{{\text{AN}}}}$. The various possible exploitation scenarios of the new infrastructure density correspond to different values of (feasible) pairs  $\left( {{{{\lambda _{{\text{UE}}}}^\prime } \mathord{\left/
 {\vphantom {{{\lambda _{{\text{UE}}}}^\prime } {{\lambda _{{\text{UE}}}}}}} \right.
 \kern-\nulldelimiterspace} {{\lambda _{{\text{UE}}}}}},{{{r_0}^\prime } \mathord{\left/
 {\vphantom {{{r_0}^\prime } {{r_0}}}} \right.
 \kern-\nulldelimiterspace} {{r_0}}}} \right)$, where ${\lambda _{{\text{UE}}}}^\prime ,{r_0}^\prime $  are the UE density and per-UE median rate supported by the densified network, respectively. As expected, there exists a tradeoff between increasing ${\lambda _{\text{UE}}}$  and  $r_0$. One criterion (others could be employed as well) for choosing the operational pair is the area capacity  of the densified network, given by  ${\lambda _{{\text{AN}}}}^\prime  \cdot {r_0}^\prime $, which, for the case considered in Figure \ref{Fig4}, slightly favors higher  ${\lambda _{{\text{UE}}}}^\prime $ and smaller  ${r_0}^\prime $.

\section{Coordination Benefits for UDNs}

The analysis in Section IV provides performance insights for a baseline UDN that operates in an uncoordinated fashion: each AN eliminates intra-cell interference by employing an OFDMA-like scheme to serve its UEs, with no action taken with regards to the generated inter-cell interference. Clearly, incorporation of network-wise resource management mechanisms that coordinate AN transmissions/interference is a natural step towards enhancing overall system performance. 

However, it is not clear whether, under the basic system model assumptions of the previous section, coordination within the UDN context can indeed provide (significant) benefits as it may be argued that the proximity gains exploited by the baseline UDN alone are sufficient. Note that this statement is true at least asymptotically, i.e., for $\tau  \to \infty$ , since the SIR becomes arbitrarily large. Coordination in an UDN involves:
\begin{itemize}
\item association of each UE with a (not necessarily closest) serving AN, taking into account channel and traffic load conditions,
\item tuning of power and/or multiple-antenna precoder for each AN-UE communication pair, utilizing the available degrees of freedom towards focusing ANs signals to intended UEs while reducing interference to/from neighbor UEs,
\item allocation of AN-UE pairs into orthogonal time-frequency resource blocks, aiming at reducing the interference among near-by users served by different ANs.
\end{itemize}

Although (optimized) network coordination is a promising technology for future UDNs, it entails two major challenges. First, the relevant coordination problems are typically highly complex large-scale optimization problems, for which finding the optimal decision with reasonable complexity becomes quite challenging. This is due to i) the vast degrees of freedom and allocation opportunities available due to increased network density, and ii) combination of both continuous and discrete optimization variables, e.g., power levels and AN-UE associations. Secondly, from an implementation perspective, network-wise coordination requires a centralized entity having access to instantaneous network channel state information. This imposes significant challenges on overhead signaling bandwidth and reporting latency requirements, which are also difficult to model and analyze. However, performance of coordination even under ideal assumptions on e.g., channel state information or complexity, is certainly of interest as small performance gains in this case will imply marginal gains or even performance loss when practical limitations are also considered.

Initial results on coordination benefits for UDNs may be found in \cite{GoSt14iswcs}, where a comprehensive optimization framework was introduced and explored under perfect channel state information and various backhaul limitation scenarios. In this paper we explore the impact of coordination on UDNs, by comparing the performance of two indicative, suboptimal/low-complexity coordination policies with the baseline approach of the previous section. In particular, the following policies are introduced \cite{GoAl13}:
\begin{enumerate}
\item ``Policy I'', which associates each UE to his/her nearest AN, performs constant power loading, and dynamically (based on instantaneous channel conditions) manages interference levels, by intelligently allocating the highest interfering AN-UE pairs to orthogonal frequencies,
\item ``Policy II'', which builds on Policy I, and, as additional performance-enhancement step, applies dynamic power loading over the AN-UE pairs.
\end{enumerate}

\begin{figure}
\centering
\resizebox{7.5 cm}{!}{\includegraphics{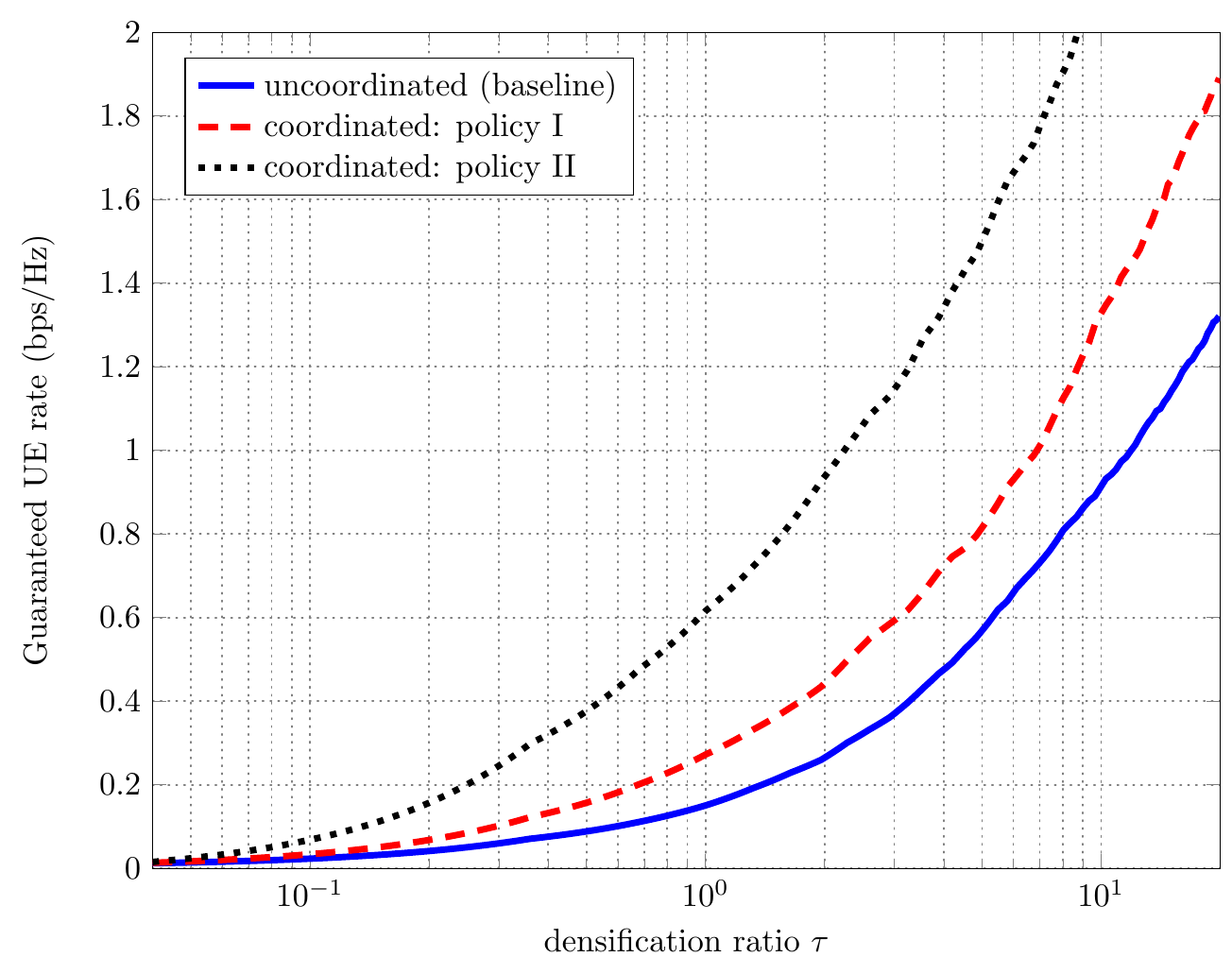}}
\caption{Guaranteed UE rate provided by baseline and coordinated UDNs (network area $1 \text{km}^2$, $\alpha=4$, maximum transmitted power per AN: $30$ dBm, noise power spectral density: $-174$ dBm/Hz).}
\label{Fig5}
\end{figure}

Figure \ref{Fig5} illustrates the minimum guaranteed user rate as a function of  $\tau$, provided by the baseline as well as the coordinated policies for a finite-area network with a variable number of randomly positioned ANs  and 50 randomly positioned UEs. Results are obtained by averaging over multiple independent realizations of the system. As observed, both coordination algorithms provide significant performance gains as compared to the uncoordinated network operation, for all densification levels. Interestingly, for values of  as high as 20, benefits of coordination are clearly visible, suggesting that proximity alone is not sufficient for practical values of AN densities. 

\begin{table}
\renewcommand{\arraystretch}{1.0}
\caption{Densification Savings Provided by Coordinated Policies Compared to Baseline}
\label{ta:Table1}
\centering
\begin{tabular}{ |p{8em}||p{5em}|p{5em}| }
\hline
\multirow{2}{10em}{Guaranteed UE Rate (bps/Hz)}  &\multicolumn{2}{|c|}{\% Savings in Densification Ratio} \\ \cline{2-3}
& Policy I & Policy II \\ \hline \hline
0.1 & 60 \% & 86 \%  \\
0.5 & 45 \%  & 82 \%  \\
1 & 41 \%  & 81 \%  \\
\hline
\end{tabular}
\end{table}

Alternatively, the coordinated policies may be viewed as a means of reducing the densification requirements for achieving certain performance levels. Indeed, Table \ref{ta:Table1} presents the savings provided by the two coordination policies in required densification ratio with respect to the baseline case, for three indicative guaranteed UE rate levels, that is 0.1, 0.5 and 1 bps/Hz. Clearly, coordination offers significant gains: applying the simpler policy I provides 40 - 60 \% infrastructure savings, whereas the more sophisticated policy II provides savings that reach 80 - 86 \%, depending on the required rate levels. As expected, these savings decrease with increasing densification levels as proximity gains become more dominant, however, they remain significant even for AN densities of an order of a magnitude greater than UE densities. These results demonstrate the potential of coordination in UDNs, although a study under realistic assumptions on channel state information and complexity is certainly necessary.

\section{Conclusions, Open Issues, and Future Challenges}
Ultra network densification is considered the main enabler for 5G wireless access. The future envisioned scenarios involve massive user density volumes and data rates per user device, challenging any evolutionary approach of cellular networks. The UDN paradigm aspires to revolutionize current cellular thinking by efficiently exploiting the advantages offered by proximal communications, however, it has to do so under a highly challenging interference environment. In this article, some of the fundamental aspects of this new network paradigm were investigated. Using an analytical framework, the scaling laws of the AN density required to support a given UE density with a given rate per UE requirement were presented, and the possibilities of exploiting the system capacity increase due to network densification were discussed. The benefits of coordination as part of UDN operation were investigated under two simple schemes where it was shown that even for relatively high densification ratios, coordination has the potential to significantly enhance performance either in terms of achievable per UE rate or minimum AN density required to support a certain rate.

Having addressed a series of fundamental issues expected to arise in the emerging UDN concept, this article aims at stimulating further research towards addressing a number of open issues. Beginning with the fundamental limits characterization, enhanced analytical studies will be needed for understanding UDNs full potential. These studies should accurately capture realistic network features not addressed before, such as space division multiple access, dynamic (in space and/or time) resource management, more elaborate coordination and cooperation capabilities among ANs, etc. Focusing on more practical concerns for UDNs, new approaches for performance modeling, taking into account backhaul overhead, over-the-air signaling, computational complexity, and cost requirements, are required for coming up with efficient and realistic network deployment strategies. Since UDN is considered the major enabler for future wireless access, it should also be studied in conjunction with other potential 5G pillars, namely enhancements in spectrum  (e.g. through large mmWave bands consideration) and spectral efficiency (e.g. massive and full-dimension MIMO), ultimately targeting the development of a comprehensive 5G performance analysis framework.

\bibliographystyle{IEEEtran}
\bibliography{IEEEabrv,ARTCOMP-bib}

\end{document}